\documentclass[a4paper,11pt]{article}
\pdfoutput=1 

\usepackage{jheppub} 

\usepackage[T1]{fontenc} 

\usepackage[latin1]{inputenc}
\usepackage[english]{babel}

\usepackage{amssymb,amsthm,cancel,hyperref,graphicx,xcolor}
\usepackage{picinpar,graphicx,xypic}
\usepackage{booktabs}
\usepackage{mathrsfs}
\usepackage{amsfonts}
\usepackage{latexsym}
\usepackage{booktabs,graphicx,hyperref,epsfig}
\usepackage{bm}
\allowdisplaybreaks
\def\bea{\begin{align}}
\def\eea{\end{align}}
\def\beq{\begin{equation}}
\def\eeq{\end{equation}}
\def\ba{\begin{eqnarray}}
\def\ea{\end{eqnarray}}
\def\be{\begin{equation}}
\def\ee{\end{equation}}

\newcommand{\Lum}{\mathcal{L}}
\newcommand{\LumM}{\boldsymbol{\mathcal{L}}}

\newcommand{\abs}[1]{\left|\,#1\,\right|}

\newcommand{\Ord}{\mathcal{O}}
\newcommand{\gsim}{\gtrsim}

\newcommand{\sss}{\scriptscriptstyle\rm}
\newcommand{\as}{\alpha_{\sss S}}

\def\({\left(}
\def\){\right)}
\def\[{\left[}
\def\]{\right]}
\def\nzscet{N_0^{\rm\scriptscriptstyle SCET}}
\def\nzdqcd{N_0^{\rm\scriptscriptstyle dQCD}}

\def\Cscet{C^{\rm\scriptscriptstyle SCET}}
\def\CscetM{{\bm C}^{\rm\scriptscriptstyle SCET}}

\def\CqcdM{{\bm C}^{\rm\scriptscriptstyle dQCD}}
\def\sigscet{\sigma^{\rm\scriptscriptstyle SCET}}

\def\sigqcd{\sigma^{\rm\scriptscriptstyle dQCD}}

\def    \hepph  #1 {{\tt hep-ph/#1}}
\def    \hepex  #1 {{\tt hep-ex/#1}}
\long\def\symbolfootnote[#1]#2{\begingroup%
\def\thefootnote{\fnsymbol{footnote}}\footnote[#1]{#2}\endgroup}

\numberwithin{equation}{section}
\title{\boldmath Resummation prescriptions and ambiguities in
  SCET vs.\ direct QCD:  Higgs production as a case study}

\author[a,1]{Marco Bonvini,\note{Current address: Rudolf Peierls Centre
    for Theoretical Physics, 1 Keble Road, University of Oxford, OX1
    3NP, Oxford, UK}}
\author[b,c]{Stefano Forte,}
\author[d]{Giovanni Ridolfi}
\author[b,1]{and Luca Rottoli}

\affiliation[a]{Deutsches Elektronen-Synchroton, DESY,\\
Notkestra{\ss}e 85, D-22603 Hamburg, Germany}
\affiliation[b]{Dipartimento di Fisica, Universit\`a di Milano\\
Via Celoria 16, I-20133 Milano, Italy}
\affiliation[c]{INFN, Sezione di Milano, Via Celoria 16, I-20133 Milano, Italy}
\affiliation[d]{Dipartimento di Fisica, Universit\`a di Genova and
INFN, Sezione di Genova,\\
Via Dodecaneso 33, I-16146 Genova, Italy}

\emailAdd{stefano.forte@mi.infn.it}
\emailAdd{marco.bonvini@desy.de}
\emailAdd{giovanni.ridolfi@ge.infn.it}
\emailAdd{luca.rottoli@physics.ox.ac.uk}

\preprint{
\begin{flushright}
DESY 14-154\\
IFUM-1033-FT
\end{flushright}
}

\abstract{We perform a comparison of soft-gluon resummation in SCET
  vs.\ direct QCD (dQCD), using Higgs boson production in gluon fusion
  as a case study, with the goal of tracing the quantitative impact of
  each source of difference between the two approaches. We show that
  saddle-point methods enable a direct quantitative comparison despite
  the fact that the scale which is resummed in the two approaches is
  not the same. As a byproduct, we put in one-to-one analytic
  correspondence various features of either approach: specifically, we
  show how the SCET method for treating the Landau pole can be
  implemented in dQCD, and how the resummation of the optimal partonic
  scale of dQCD can be implemented in SCET.  We conclude that the main
  quantitative difference comes from power-suppressed subleading
  contributions, which could in fact be freely tuned in either
  approach, and not really characteristic of either.  This conclusion
  holds for Higgs production in gluon fusion, but it is in fact
  generic for processes with similar kinematics. For Higgs production,
  everything else being equal, SCET resummation at NNLL in the
  Becher-Neubert implementation leads to essentially no enhancement of
  the NNLO cross-section, unlike dQCD in the standard implementation
  of Catani et al..  }

\begin{document} 
\maketitle
\flushbottom

\section{Sudakov resummation: advantages and ambiguities}
\label{sec:intro}

The current standard for the computation of hard processes at the LHC,
such as Higgs production in gluon
fusion~\cite{Dittmaier:2011ti,deFlorian:2012yg}, 
is to improve fixed
order computations through the inclusion of soft-gluon
resummation. This is often advantageous even for processes which are
far from threshold, where the contributions which are resummed would
become of order one, because it may provide a good approximation to
the first few missing higher order corrections: this is indeed what
happens for Higgs production in gluon
fusion~\cite{Ball:2013bra,Bonvini:2014jma}. However, especially when
dealing with processes which are far from threshold, alternative
implementations of threshold resummation which differ by subleading
terms lead to predictions which might differ by an amount which is
comparable to the effect of the resummation itself. Comparison of
different resummation prescriptions and methodologies may then be of
considerable phenomenological interest.

A particularly relevant instance of this situation is the comparison
of results obtained when resummation is performed through the direct
use of perturbative QCD (direct QCD, or dQCD, henceforth), or using an
effective field theory (soft-collinear effective theory, or SCET,
henceforth).  For example, in the case of Higgs in gluon
fusion~\cite{Ahrens:2008qu,Ahrens:2008nc,Ahrens:2010rs} the effect of
resummation can be by a factor two larger according to whether
resummation is performed using SCET or
dQCD~\cite{Ahrens:2008nc}. Whereas there are indications that this is
likely related to power-suppressed corrections~\cite{Ahrens:2008nc},
it would be highly desirable to have a detailed quantitative
understanding of the relation between the two approaches.

In recent years there has been an increasing interest in a deeper
understanding of similarities and differences between the SCET and
dQCD approach to resummation. Whereas in the SCET papers
refs.~\cite{Becher:2006mr,Becher:2007ty,Ahrens:2008nc} a first
exploration of the relationship between the two approaches was
performed, in refs.~\cite{Bonvini:2012az,Bonvini:2013td} expressions
relating SCET and dQCD resummation for various choices of soft scale
were presented in analytic form, and finally, in
refs.~\cite{Sterman:2013nya,Almeida:2014uva}, the relationship between
the two formalisms was traced to the way they handle the underlying
soft-gluon factorization.

The full machinery which is necessary for a detailed quantitative
comparison is thus available: the comparison will be the subject of
the present paper.  The goal of this work is complementary to that of
refs.~\cite{Bonvini:2012az,Bonvini:2013td,Sterman:2013nya,Almeida:2014uva}:
there, the aim was to show the equivalence of SCET and dQCD
resummation by classifying terms which are resummed in either approach
in terms of logarithmic accuracy. Thus, in particular, in
ref.~\cite{Bonvini:2012az,Bonvini:2013td} it was shown that dQCD and
SCET resummed expressions differ by subleading logarithms of the
hadronic scale, under suitable assumptions on the parton
luminosity. Here, we will trace each ingredient which enters either
resummed expression, retaining all differences, both logarithmically-
and power-suppressed, and assess their impact.  While
refs.~\cite{Bonvini:2012az,Bonvini:2013td} focused on the invariant
mass distribution of Drell-Yan pairs, and ref.~\cite{Almeida:2014uva}
on event shapes, here we will consider specifically Higgs production
in gluon fusion. We will however always display results for a wide
range of values of the Higgs mass: much wider, in fact, than the
acceptable physical Higgs mass range. The goal is to study the
qualitative behaviour in a wide region which may be of interest for
other processes (such as Drell-Yan), while providing quantitatively
precise prediction for physical Higgs production.

Soft-gluon resummation for hadronic processes in dQCD, as derived in
refs.~\cite{Catani:1989ne,Sterman:1986aj,Contopanagos:1996nh,Forte:2002ni},
is almost universally performed using the formalism of
ref.~\cite{Catani:1996yz}, which we will take as a reference for the
dQCD approach. On the other hand, the SCET approach comes in many
flavors, which differ not only in technical details but, more
importantly, in the choice of scale which is being resummed. Here, we
refer specifically to the SCET approach of
refs.~\cite{Becher:2006nr,Becher:2006mr,Becher:2007ty,Ahrens:2008qu},
which builds upon the derivation from
SCET~\cite{Bauer:2000ew,Bauer:2000yr,Bauer:2001ct,Bauer:2001yt,Bauer:2002nz}
of soft-gluon
resummation~\cite{Manohar:2003vb,Pecjak:2005uh,Chay:2005rz,Idilbi:2005ky}. This
choice is not only motivated by the widespread use of this approach,
but also by the fact that, because it is based on  momentum space
(as opposed to
Mellin space), and a hadronic (instead of partonic)
resummed scale choice, this version of SCET resummation is in some
sense maximally different from  dQCD, and thus it will
allow us to explore all facets of the difference between SCET and
dQCD. Henceforth, we will refer to these two approaches as dQCD and
SCET, for short.

The starting point of our paper is
a  derivation,  presented in Sect.~\ref{sec:analytical}, of 
the relation between resummed
expressions in dQCD and SCET which shows how starting with
the dQCD expression one can arrive at the SCET result through a
step-by-step modification of the starting expression. This derivation
has the interesting byproduct of showing how some features of the SCET
approach (such as the removal of the Landau pole or the use of a
hadronic scale) could be implemented in a dQCD approach without having
to use SCET, and by simply manipulating the dQCD expression. It is also
more powerful than the master
formula relating SCET to dQCD of
refs.~\cite{Bonvini:2012az,Bonvini:2013td}: whereas that relation only
allowed for a classification of the logarithmic order of the
difference, this new derivation keeps track of all individual
contributions which differ in the two approaches.

Because in the SCET approach considered here it is a hadronic scale
which is being resummed, and not a partonic scale as in the dQCD
approach, it may seem that a comparison can only be made at the
hadronic level, and this indeed was the point of view taken in
refs.~\cite{Bonvini:2012az,Bonvini:2013td,Sterman:2013nya,Almeida:2014uva},
where the dependence of results on the parton luminosity was also
discussed.  However, in Sect.~\ref{sec:saddle} we show that a
comparison at the level of partonic cross-sections can be done by
means of a saddle-point method. This also allows us to assess the
quantitative impact of a first obvious source of difference between
the dQCD and SCET approaches, namely, that because of the choice of
resumming a hadronic scale the resummed SCET expression for the
hadronic cross-section does not respect the standard factorized form
of QCD expressions, in that it does not have the form of a convolution
between a parton luminosity and a partonic cross section, which
reduces to an ordinary product upon taking a Mellin transform.

With this problem out of the way, in Sect.~\ref{sec:pheno} we can then
compare the partonic cross-section as obtained in either approach. We
will in particular discuss separately two classes of contributions,
which correspond to individual steps in the procedure previously
discussed in Sect.~\ref{sec:analytical} which takes from the dQCD to
the SCET expression: subleading logarithmic terms, which arise as a
consequence of the shift from a partonic to a hadronic resummed scale,
and the specific choice of power-suppressed terms which characterizes
the standard SCET expression in comparison to the standard dQCD
expression.

Specifically we shall show that, at least with the scale
choice of
refs.~\cite{Becher:2006nr,Becher:2006mr,Becher:2007ty,Ahrens:2008qu},
power-suppressed terms 
are by far the dominant source of difference. For Higgs production in
gluon fusion, this difference turns out to be comparable to the effect
of the resummation itself: indeed,  
the NNLO+NNLL result for gluon fusion, which in
ref.~\cite{deFlorian:2012yg} using a dQCD approach
is found to be of order of 10\%, is reduced to a negligible enhancement
at the percent level if everything else is kept equal, but the SCET
approach of ref.~\cite{Ahrens:2008nc,Ahrens:2010rs} is used instead,
with the difference entirely due to power suppressed terms. 

This is an interesting
conclusion, because such terms are, in fact, unrelated to the choice
of resumming via SCET or dQCD; rather, they may be freely tuned in
either approach. So the difference is large indeed, but not intrinsically
related to the use of dQCD vs. SCET, and specifically not to the
treatment of the Landau pole, or the choice of resumming a hadronic instead
of partonic scale. Rather, this is yet another instance of 
the fact that large ambiguities may come from power
suppressed terms when resummation is used to improve perturbative
predictions away from the threshold region, as already pointed out by
us some
time ago~\cite{Bonvini:2010tp} in the context of Drell-Yan
production. In the specific context of gluon fusion, it has been known
for a long time~\cite{Kramer:1996iq} that 
some choices of power suppressed terms lead to an improved agreement
of the truncation of resummed results with known fixed-order results.
More recently, the large size of power-suppressed ambiguities to the
soft approximation to Higgs production in gluon fusion computed at
N$^3$LO was pointed out in ref.~\cite{Anastasiou:2014vaa}, and a
general systematic way of optimizing such corrections was suggested 
in  ref.~\cite{Ball:2013bra}. We shall briefly comment
on the implications of this in our concluding Sect.~\ref{sec:conclusions}.

\section{Resummation: from  dQCD to SCET}
\label{sec:analytical}

In this Section we present a step-by-step argument which takes from
the dQCD to the SCET form of the resummed cross-section.  
We will give our argument with
specific reference to Higgs boson production in gluon fusion, up to
NNLL accuracy.

We define the dimensionless cross-section $\sigma$ in terms of the physical
cross section $\sigma_{\rm Higgs}$ as
\be 
\sigma(\tau,m_H^2)=\frac{1}{\tau \sigma_0}
\sigma_\textrm{Higgs}(\tau,m_H^2)
=\int_\tau^1 \frac{dz}{z}\, C(z,m_H^2) \Lum\left(\frac{\tau}{z}
\right),
\label{sigmahad}
\ee
where $\sigma_0$ is the leading-order partonic cross section for the 
process $gg\to H$, $m_H$ is the Higgs mass,
$\tau=\frac{m_H^2}{s}$ and $s$ is the hadronic center-of-mass energy squared,
$\Lum$ is the gluon luminosity, and $C$ a dimensionless
coefficient function related to the partonic cross section. 
For simplicity, we do not show the dependence
on factorization and renormalization scales, which we choose
to be equal to $m_H$ throughout the paper.

The quantity in eq.~\eqref{sigmahad} factorizes upon
Mellin transformation with respect to $\tau$:
\be\label{eq:mellin}
\boldsymbol{\sigma}(N,m_H^2) = \int_0^1 d\tau\, \tau^{N-1} \sigma(\tau,m_H^2) 
= {\bm C}(N,m_H^2)\LumM(N)
\ee
where
\be
{\bm C}(N,m_H^2)=\int_0^1dz\,z^{N-1}C(z,m_H^2);
\qquad \LumM(N)=\int_0^1dz\,z^{N-1}\Lum(z).
\ee
Because of the importance of keeping track of the difference between
Mellin space and momentum space, we refrain from the common abuse of
notation (of which we were specifically guilty in
refs.~\cite{Bonvini:2012az,Bonvini:2013td}) whereby the same notation
is used for a function and its
Mellin transform; rather, we will follow 
the convention that a function and its Mellin transform 
are denoted with the same symbol, but with the Mellin transform in boldface.

The resummed $N$-space coefficient function in dQCD 
can be written as
\be
\label{eq:cqcdhat}
\CqcdM(N,m_H^2) =
\hat g_0 (\as(m_H^2)) \exp \hat S^{\rm\scriptscriptstyle dQCD}
\left( m_H^2, \frac{m_H^2}{\bar N^2} \right)
\ee
where $\bar N = N e^{\gamma_E}$ and 
\be\label{sdqcd}
\hat S^{\rm\scriptscriptstyle dQCD}\left(\mu_1^2,\mu_2^2\right) =
\int_{\mu_1^2}^{\mu_2^2} \frac{d\mu^2}{\mu^2} 
\left[A(\as(\mu^2)) \ln \frac{m_H^2}{\mu^2 \bar N^2}+
\hat D(\as(\mu^2)) \right].
\ee
For later convenience, the two arguments of
$\hat S^{\rm\scriptscriptstyle dQCD}$ have been chosen to be the two integration
bounds. 

The functions $\hat g_0(\as), A(\as)$ and $\hat D(\as)$ admit Taylor
expansions in $\as$; the order at which the expansions are truncated
defines the logarithmic accuracy of the resummation, as indicated in
table~\ref{tab:count}.
\begin{table*}[t]
\begin{center}
\begin{tabular}{lcccccc}
\toprule
&(dQCD)  & $A(\as)$ & $\hat D(\as)$ & $\hat g_0(\as)$ & & $\as^n \ln^kN$\\
&(SCET) & $\Gamma_{\rm cusp}(\as)$ & $\gamma_W(\as)$ & $H$, $\tilde s_{\rm Higgs}$ &
&  $\as^n \ln^k(\mu_s/M)$\\
\midrule
  LL    & & 1-loop & --- & tree-level & & $k= 2n$ \\
\cmidrule(){1-7}
  NLL*  & & 2-loop & 1-loop & tree-level & & $2n-1\le k\le 2n$ \\
  NLL   & & 2-loop & 1-loop & 1-loop & & $2n-2\le k\le 2n$ \\
\cmidrule(){1-7}
  NNLL* & & 3-loop & 2-loop & 1-loop & & $2n-3\le k\le 2n$ \\
  NNLL  & & 3-loop & 2-loop & 2-loop & & $2n-4\le k\le 2n$\\
\bottomrule
\end{tabular}
\caption{Orders of logarithmic approximations in resummed
  computations. The central three columns of the table give the order
  at which the functions which enter respectively the dQCD or SCET
  resummed expressions, as listed above the table, must be computed in
  order to achieve the accuracy called with the name given in the
  first column, and corresponding to the inclusion in the coefficient
  function of terms as given in the last column.}
\label{tab:count}
\end{center}
\end{table*}
 In particular, to NNLL accuracy,
\begin{align}\label{A2NNLL}
&A(\as)=\frac{A_1}{4}\as+\frac{A_2}{16}\as^2+\frac{A_3}{64}\as^3
\\
&\hat D(\as)=\hat D_2\as^2.
\label{D2NNLL}
\end{align}
Note that there is an ambiguity in the way the expression
eq.~\eqref{eq:cqcdhat} is defined: specifically, one may 
rewrite eq.~\eqref{eq:cqcdhat} as 
\begin{align}
&
\CqcdM(N,m_H^2) =
g_0(\as(m_H^2)) \exp S^{\rm\scriptscriptstyle dQCD}
(\bar \alpha \ell,\bar\alpha) 
\label{eq:CresgS}
\\
&\bar \alpha \equiv 2 \beta_0 \as, \qquad \ell \equiv \ln \frac{1}{N}
\end{align}
where
\begin{align}
&g_0(\as) = 1 + \sum_{j=1}^\infty g_{0j} \as^j
\\
&S^{\rm\scriptscriptstyle dQCD}
(\bar \alpha \ell,\bar\alpha) 
=\frac{1}{\bar \alpha} g_1(\bar \alpha \ell) +g_2(\bar \alpha \ell) 
+ \bar \alpha g_3 (\bar \alpha \ell) 
+ \bar \alpha^2 g_4(\bar \alpha\ell) + \ldots.
\end{align}
This differs from eq.~\eqref{eq:cqcdhat} because now all constant 
(i.e., $N$-independent) terms are
included in $g_0(\as)$, while in eq.~\eqref{eq:cqcdhat} some constant
terms were exponentiated; of course, the difference is logarithmically
subleading. In the sequel, we will use eq.~\eqref{eq:CresgS} for
numerical implementations.

We want to relate the dQCD expression
eqs.~(\ref{eq:cqcdhat}--\ref{sdqcd}) to its SCET counterpart. The latter,
in the SCET approach of
refs.~\cite{Becher:2006nr,Becher:2006mr,Becher:2007ty,Ahrens:2008nc}
is given directly in the space of the physical variable $z$, and
it can be written as
\be
\label{eq:resummdyscetonescale} 
\Cscet(z,m_H^2,\mu_s^2) =
H(m_H^2) U(m_H^2,\mu_s^2) S(z,m_H^2,\mu_s^2),
\ee 
(the SCET result is actually more generally
expressed as a function of other
energy scales, which here we all take  to be equal to the hard scale $m_H$ 
for simplicity, as this does not affect our arguments).
The soft function is given by
\be
\label{softfunction}
S(z, m_H^2, \mu_s^2) = \tilde s_\textrm{Higgs}
\left(\ln\frac{m_H^2}{\mu_s^2}+\frac{\partial}{\partial\eta},\mu_s\right)
\frac{1}{1-z} \left(\frac{1-z}{\sqrt{z}}\right)^{2 \eta} \frac{e^{-2
    \gamma_E \eta}}{\Gamma(2 \eta)} 
\ee
where 
\be 
\eta =\int_{m_H^2}^{\mu_s^2} \frac{d\mu^2}{\mu^2}\,A(\as(\mu^2)).
\label{etadef}
\ee
Finally,
\be
U(m_H^2,\mu_s^2)=\exp\int_{m_H^2}^{\mu_s^2} \frac{d\mu^2}{\mu^2}\,
\left[A(\as(\mu^2)) \ln
  \frac{m_H^2}{\mu^2}+\gamma_W(\as(\mu^2)) \right]
\ee 
with $\gamma_W(\as)=\gamma_{W,2}\as^2$ to NNLL accuracy.
The hard function $H(m_H^2)$ admits an expansion in powers of
$\as(m_H^2)$, and $\tilde s_\textrm{Higgs}(L,\mu_s^2)$ admits an
expansion in powers of $\as(\mu_s^2)$. The logarithmic accuracy of the
resummation is fixed by the order at which these expansion are
truncated, as shown in table~\ref{tab:count}. Note that 
there is a slight ambiguity in the way the resummed expression
eq.~\eqref{eq:resummdyscetonescale} is defined: in particular, once
the various factor on the right-hand side are included to any desired
order, the interference between them generates higher order terms: for
example, if at NLL both $H$ and  $\tilde s_\textrm{Higgs}$ are
included up to order $\as$, their product will clearly include
terms $\Ord(\as^2)$. In this
paper, specifically for all numerical implementations to be discussed
in  the sequel, we will consistently drop all higher-order
interference terms, because  this is the prescription
used in the implementation of the results of
refs.~\cite{Ahrens:2008qu,Ahrens:2008nc,Ahrens:2010rs} in the {\tt
  RGHiggs} code~\cite{rghiggs}, as we have explicitly checked. 
This choice spoils the factorized form of the SCET
result, eq.~(\ref{eq:resummdyscetonescale}),
 though it simplifies the matching to fixed
order. 

\begin{figure}[t]
\centering
{\includegraphics[width=.49\columnwidth]
{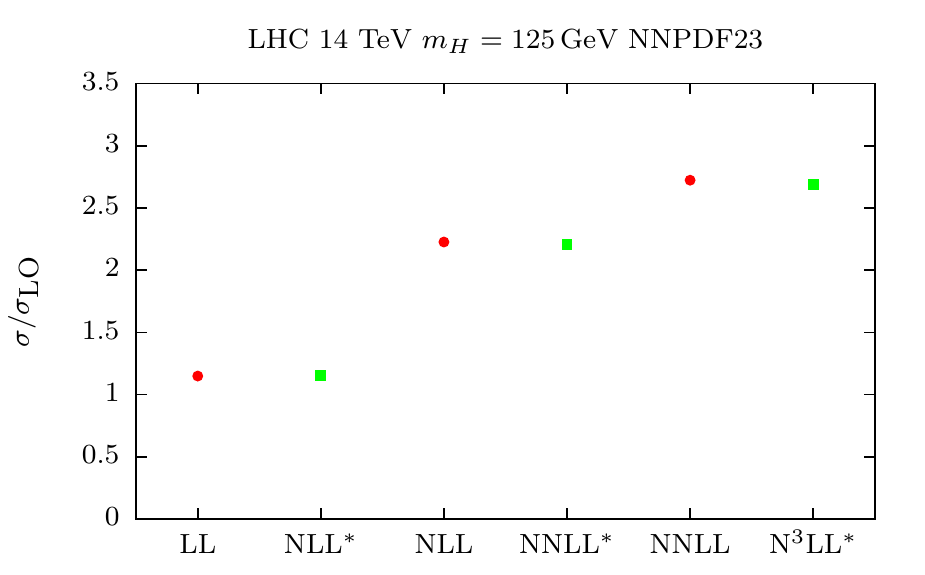}\includegraphics[width=.49\columnwidth]
{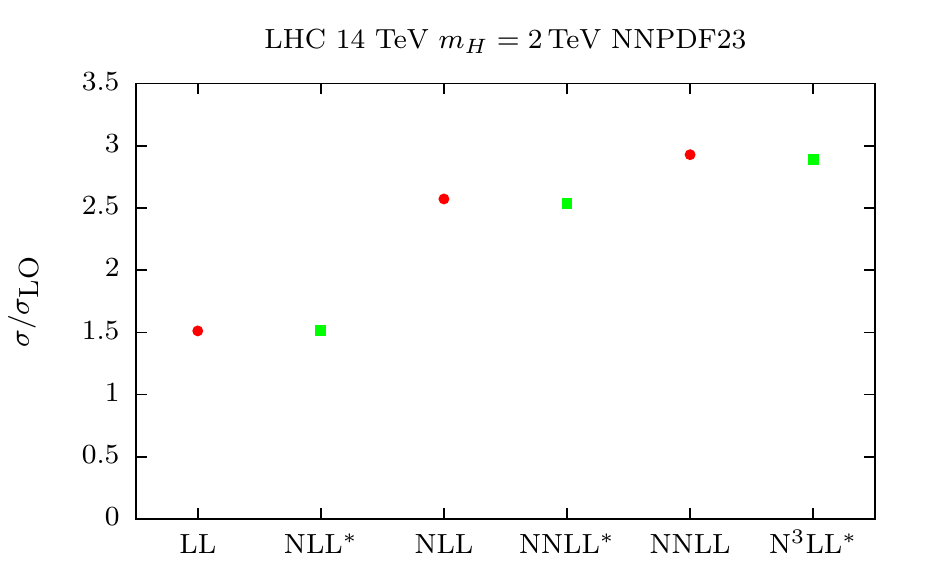}}
\caption{The cross-section for production of a Higgs-like particle of
  mass $m_H=125$~GeV (left) or  $m_H=2$~TeV (right) in gluon fusion at the LHC~14~TeV computed using the
  SCET expression at the various resummed orders of
  table~\ref{tab:count}. Results have been obtained using NNPDF2.3 NNLO
  PDFs with $\alpha_s(M_Z^2)=0.118$ and $\mu_R=\mu_F=m_H$. 
}\label{fig:starvsnonstar}
\end{figure}
Note that in 
refs.~\cite{Becher:2006nr,Becher:2006mr,Becher:2007ty,Ahrens:2008qu}
all results were presented in the starred (NLL*, NNLL* and so forth)
approximation (and, somewhat confusingly, referred
to as NLL and NNLL). The way the result of
refs.~\cite{Becher:2006nr,Becher:2006mr,Becher:2007ty,Ahrens:2008qu}  
 could be
upgraded to the more accurate NNLL approximation was presented in
ref.~\cite{Bonvini:2013td} (SCET results corresponding to either
starred or unstarred accuracy were considered e.g.\ in ref.~\cite{Berger:2010xi}
in the context of the resummation of jet vetoes).

For reference, in  figure~\ref{fig:starvsnonstar} we compare the 
 total cross section for Higgs production in gluon fusion at the
LHC~14~TeV, determined at the various resummed orders of
table~\ref{tab:count}, both using the starred result of
Refs.~\cite{Becher:2006nr,Becher:2006mr,Becher:2007ty,Ahrens:2008qu}
and its unstarred upgrade  
of Ref.~\cite{Bonvini:2013td}, for the physical value of the Higgs
mass and for a very high-mass scalar, much closer to threshold (all other settings are the same used in
all other plots in the sequel of the paper). 
While only the SCET result is shown in the
plot, the dQCD result follows a similar pattern. It is clear that
upgrading from the nonstarred to the starred result has a very
substantial effect, while going up from  the starred result at
one order to the unstarred result at the subsequent order only has a very
minor impact.
Note also that it is natural to match the generic fixed order N$^k$LO
to either N$^k$LL or N$^{k+1}$LL*, as done respectively in, e.g.,
Ref.~\cite{deFlorian:2012yg} and Refs.~\cite{Becher:2006nr,Becher:2006mr,Becher:2007ty,Ahrens:2008qu};
figure~\ref{fig:starvsnonstar} shows that the extra effort needed in order to upgrade
from N$^k$LL to N$^{k+1}$LL* is unnecessary.
Consequently we will  henceforth
only consider and compare the unstarred (NLL, NNLL etc.) results.

We now start from the dQCD expression
eqs.~(\ref{eq:cqcdhat}--\ref{sdqcd}).
It is well known that $\CqcdM(N,m_H^2)$,
considered as a function of the complex variable $N$,
has a branch cut on the positive real axis for $\bar N\ge N_L$, where
$N_L$ is the position of the Landau pole of $\as(m_H^2/\bar N^2)$:
\be
1-2\beta_0\as(m_H^2)\ln N_L=0.
\ee
As a consequence, $\CqcdM(N,m_H^2)$ does not have an inverse Mellin
transform.  This is an effect of resummation: indeed, any finite-order
truncation of the expansion of $\CqcdM(N,m_H^2)$ in powers of
$\as(m_H^2)$ does have a Mellin inverse~\cite{Forte:2006mi}.

We now show that a possible solution of the Landau pole problem,
suggested by the structure of the SCET result, leads us directly to
the SCET expression eq.~\eqref{eq:resummdyscetonescale} in a few
simple steps. We first note that  
the exponent in the dQCD expression can be rewritten
\be
\hat S^{\rm\scriptscriptstyle dQCD}\left(m_H^2,\frac{m_H^2}{\bar N^2} \right)=
\hat S^{\rm\scriptscriptstyle dQCD}\left(m_H^2,\mu_s^2\right)+
\hat S^{\rm\scriptscriptstyle dQCD}\left(\mu_s^2,\frac{m_H^2}{\bar N^2} \right),
\label{Sqcdsplit}
\ee
where $\mu_s$ is an arbitrary energy scale. 
Now assume that we can choose $\mu_s$
so that $\as(\mu_s^2)$ is in the perturbative domain, and
at the same time
\be\label{eq:logcond}
\abs{\ln\frac{\mu_s^2}{m_H^2}}\gg
\abs{\ln\frac{m_H^2}{\mu_s^2\bar N^2}}
\ee
for all values of $N$ in the relevant range (whether such an energy
scale actually exists, and how it can be determined in the context of
dQCD, are issues that will be discussed in the next section).  This
means that $\mu_s$ is chosen to be of the same order of
$\frac{m_H^2}{\bar N^2}$, the scale which is being resummed in the
dQCD approach.

In such case, the first term in eq.~\eqref{Sqcdsplit} contains the
resummation, while the second term can be expanded in powers if
$\as(\mu_s^2)$: it is on the same footing as the function $\hat
g_0(\as)$ of eq.~\eqref{eq:cqcdhat}, because now
$\ln\frac{m_H^2}{\mu_s^2\bar N^2}$ does not count as a large log
(the large log is instead $\ln\frac{\mu_s^2}{m_H^2}$), and
thus to NNLL (see table~\ref{tab:count}) it must be only kept up to
$\Ord(\as^2)$, neglecting $\Ord(\as^3)$ terms. We find
\begin{align}\label{eq:dqcdtf}
\exp\hat S^{\rm\scriptscriptstyle dQCD}\left(\mu_s^2,\frac{m_H^2}{\bar N^2} \right)
&=\exp\int_{\mu_s^2}^{\frac{m_H^2}{\bar N^2}} \frac{d\mu^2}{\mu^2} 
\left[A(\as(\mu^2)) \ln \frac{m_H^2}{\mu^2 \bar N^2}+
\hat D(\as(\mu^2)) \right]
\nonumber\\
&=F(L,\mu_s)
\end{align}
where
\be\label{eq:fexp}
F(L,\mu_s)=1+\as(\mu_s^2)\frac{A_1 L^2}{8}
+\as^2(\mu_s^2)
\left[\frac{A_1^2L^4}{128}-\frac{\beta_0 A_1 L^3}{24}
+\frac{A_2L^2}{32}+\hat D_2 L\right]+\Ord(\as^3)
\ee
and
\be
L=\ln\frac{m_H^2}{\mu_s^2\bar N^2}.
\ee
Thus, we obtain a $\mu_s$-dependent version of the dQCD resummed cross
section, which reads
\begin{align}
\CqcdM(N,m_H^2,\mu_s^2)&=\hat g_0(\as(m_H^2))F(L,\mu_s)
\nonumber\\
\times&\exp\int_{m_H^2}^{\mu_s^2} \frac{d\mu^2}{\mu^2} 
\left[A(\as(\mu^2)) \ln \frac{m_H^2}{\mu^2\bar N^2}+\hat D(\as(\mu^2)) \right].
\label{Cqcdmus1}
\end{align}
Because we are keeping only a finite number of terms in the expansion
of $F(L,\mu_s)$, the inverse Mellin transform is now well defined. We
have thus arrived at a form of the dQCD result which is free of the
Landau pole problem, in analogy to the SCET expression.  Note that
while up to NNLL $A(\as)$ eq.~\eqref{D2NNLL} includes contributions up
to $\Ord(\as^3)$, only the $A_1$ and $A_2$ terms 
contribute to $F(L,\mu_s)$
eq.~\eqref{eq:fexp}. This observation will help us in understanding
the relation of the result found here with that of
ref.~\cite{Bonvini:2013td}.

We can now show that the form eq.~\eqref{Cqcdmus1} of the dQCD result
coincides with the SCET result up to subleading terms. First, we note
that the functions $F(L,\mu_s)$ and $\tilde s_{\rm Higgs}(L,\mu_s)$
are closely related. Indeed, using the explicit expressions of the
coefficients of the expansion of $\tilde s_{\rm Higgs}(L,\mu_s)$ in
powers of $\as$,
\be\label{stildef}
\tilde s_{\rm Higgs}(L,\mu_s)\exp\left(-\frac{\zeta_2 C_A}{2\pi}
\as(\mu_s^2)\right) 
=F(L,\mu_s)+k\as^2(\mu_s^2)+\Ord(\as^3),
\ee
where $k$ is a numerical constant.
Recalling that $F(0,\mu_s)=1$, we find
\be\label{tildetof}
\frac{\tilde s_{\rm Higgs}(L,\mu_s)}{\tilde s_{\rm Higgs}(0,m_H)}
\exp\left(-\frac{\zeta_2 C_A}{2\pi}\(\as(\mu_s^2)-\as(m_H^2)\)\right) 
=F(L,\mu_s)+\Ord(\as^3).
\ee
Also,
\be\label{gto H}
\hat g_0(\as(m_H^2))=H(m_H^2)\tilde s(0,m_H)\left[1+\Ord(\as^3)\right].
\ee
Finally, the first term in eq.~\eqref{Sqcdsplit} can be written as
\begin{align}\label{nfactor}
\hat S^{\rm\scriptscriptstyle dQCD}\left(m_H^2,\mu_s^2\right) &=
\int_{m_H^2}^{\mu_s^2} \frac{d\mu^2}{\mu^2} 
\left[A(\as(\mu^2)) \ln \frac{m_H^2}{\mu^2 \bar N^2}+
\hat D(\as(\mu^2)) \right]
\nonumber\\
&=\int_{m_H^2}^{\mu_s^2} \frac{d\mu^2}{\mu^2} 
\left[A(\as(\mu^2)) \ln \frac{m_H^2}{\mu^2}+
\hat D(\as(\mu^2)) \right]+\ln\bar N^{-2\eta},
\end{align}
where $\eta$ is defined in eq.~\eqref{etadef}. 

Substituting eqs.~\eqref{tildetof}--\eqref{nfactor} in the Landau-pole
free form of the dQCD result, eq.~\eqref{Cqcdmus1}, we get
\begin{align}
\CqcdM(N,m_H^2,\mu_s^2)&=H(m_H^2)\tilde s_{\rm Higgs}(L,\mu_s)\bar N^{-2\eta}
\exp\left(\frac{\zeta_2 C_A}{2\pi}(\as(m_H^2)-\as(\mu_s^2))\right)
\nonumber\\
\times&\exp\int_{m_H^2}^{\mu_s^2} \frac{d\mu^2}{\mu^2} 
\left[A(\as(\mu^2)) \ln \frac{m_H^2}{\mu^2}
+\hat D(\as(\mu^2)) \right]\left(1+\Ord(\as^3)\right).
\end{align}
This can be brought in the same form as the SCET result by noting that
\be
\int_{m_H^2}^{\mu_s^2}\frac{d\mu^2}{\mu^2} \hat D(\as(\mu^2))
=\hat D_2\int_{m_H^2}^{\mu_s^2} \frac{d\mu^2}{\mu^2}\as^2(\mu^2)
=\frac{\hat D_2}{\beta_0}(\as(m_H^2)-\as(\mu_s^2))
\ee
to NNLL accuracy, and that
\be
\hat D_2+\frac{\zeta_2 C_A}{2\pi}\beta_0=\gamma_{W,2}.
\ee
Hence
\begin{align}
\CqcdM(N,m_H^2,\mu_s^2)&=H(m_H^2)\tilde s_{\rm Higgs}(L,\mu_s)\bar N^{-2\eta}
\exp\int_{m_H^2}^{\mu_s^2} \frac{d\mu^2}{\mu^2} 
\left[A(\as(\mu^2)) \ln \frac{m_H^2}{\mu^2}+\gamma_W(\as(\mu^2)) \right]
\nonumber\\
&=H(m_H^2)U(m_H^2,\mu_s^2)\tilde s_{\rm Higgs}(L,\mu_s)\bar N^{-2\eta}
\left(1+\Ord(\as^3)\right),
\label{Cqcdmus}
\end{align}
which is the Landau-pole-free form of the dQCD result, written in SCET
factorized form.
For future convenience, we define the ratio of the starting dQCD
expression eq.~(\ref{eq:cqcdhat}) to this intermediate result
(eq.~\eqref{Cqcdmus}):
\be
\label{eq:cr1def}
{\bm C}^{(1)}_r(N,m_H^2, \mu_s^2)
\equiv\frac{\CqcdM(N,m_H^2)}{H(m_H^2)U(m_H^2,\mu_s^2)
\tilde s_\textrm{Higgs}\left(L,\mu_s\right)
\bar N^{-2 \eta}}.
\ee

It is now immediate to show that the result  eq.~\eqref{Cqcdmus} coincides
with the Mellin transform
of the SCET coefficient function, up to power-suppressed terms.
Indeed,
\be\label{eq:cscetlargenM}
\CscetM(N,m_H^2,\mu_s^2)=H(m_H^2)U(m_H^2,\mu_s^2){\bm S}(N,m_H^2,\mu_s^2),
\ee
where
\be
\label{eq:exactmt}
{\bm S}(N,m_H^2,\mu_s^2) = \tilde s_\textrm{Higgs} 
\left(\ln\frac{m_H^2}{\mu_s^2} +\frac{\partial}{\partial \eta},\mu_s\right)
\frac{\Gamma(N-\eta)}{\Gamma(N+\eta)}e^{-2 \gamma_E \eta}
\ee
for fixed (i.e., $z$-independent) $\mu_s$. But up to power-suppressed terms
\begin{align}
{\bm S}(N,m_H^2,\mu_s^2) &=
\tilde s_\textrm{Higgs}\left(\ln\frac{m_H^2}{\mu_s^2} 
+\frac{\partial}{\partial \eta}, \mu_s \right)
\bar N^{-2 \eta}+\Ord \left(\frac{1}{N} \right)
\nonumber\\
&=
\tilde s_\textrm{Higgs}\left(\ln\frac{m_H^2}{\mu_s^2\bar N^2},\mu_s\right)
\bar N^{-2 \eta}\left(1+\Ord \left(\frac{1}{N} \right)\right).
\label{SoftLogN}
\end{align}
Thus, up to terms suppressed in the large-$N$ limit,
\be
\CscetM(N,m_H^2,\mu_s^2)=H(m_H^2)U(m_H^2,\mu_s^2)
\tilde s_\textrm{Higgs}\left(L,\mu_s\right)
\bar N^{-2 \eta}\left(1+\Ord \left(\frac{1}{N} \right)\right),
\ee
which is what we set out to prove. 
We also define the ratio of the starting dQCD
expression eq.~(\ref{eq:cqcdhat}) to the exact Mellin transform
eq.~(\ref{eq:cscetlargenM}) of the SCET expression eq.~(\ref{eq:resummdyscetonescale}):
\be\label{eq:cr2def}
{\bm C}^{(2)}_r(N,m_H^2, \mu_s^2)=\frac{\CqcdM(N,m_H^2)}{\CscetM(N,m_H^2,\mu_s^2)}.
\ee

The steps leading from the Landau-pole-free dQCD expression
eq.~(\ref{Cqcdmus})
to the standard SCET expression  eq.~(\ref{eq:resummdyscetonescale}) 
can also be traced
in  $z$ space. Indeed,
the $z$-space form of the expression  eq.~\eqref{Cqcdmus} is
obtained by inverse Mellin transformation, and it is given by 
\be\label{eq:scetlarget}
S(z, M^2, \mu_s^2) = \tilde s_\textrm{Higgs}
\left(\ln\frac{M^2}{\mu_s^2}+\frac{\partial}{\partial\eta},\mu_s\right)
(-\ln z)^{-1+2 \eta} \frac{e^{-2 \gamma_E \eta}}{\Gamma(2 \eta)}.
\ee 
Noting that
\be \label{eq:powcorr}
(-\ln z)^{-1+2 \eta}=
\frac{1}{1-z} \left(\frac{1-z}{\sqrt{z}}\right)^{2 \eta}
\left[ 1 +\Ord\left( 1-z\right)\right]
\ee 
it follows that  eq.~(\ref{eq:scetlarget}) differs from
the starting SCET expression of the soft function
eq.~(\ref{softfunction}) by terms which are suppressed by positive
powers of $1-z$. It is interesting to observe that 
\be \label{eq:leadpowcorr}
(-\ln z)^{-1+2 \eta}=
\frac{\sqrt{z}}{1-z} \left(\frac{1-z}{\sqrt{z}}\right)^{2 \eta}
\left[ 1 +\Ord\left(( 1-z)^2\right)\right],
\ee
i.e., the leading power-suppressed difference can be in fact expressed
as a $\sqrt{z}$ prefactor, with further power suppressed differences
only arising at $\Ord\left(( 1-z)^2\right)$. It was pointed out in
  Ref.~\cite{Ball:2013bra} (see in particular the discussion of
  eq.~(2.43) of this reference) that this $\sqrt{z}$ prefactor has a
  non-negligible impact. While the reader is referred to
  Refs.~\cite{Ball:2013bra,Kramer:1996iq} for a discussion of the motivation and
  quantitative impact of this factor, we recall here that its
  effect  goes in the
  same direction as the so-called collinear improvement  of the
  resummed result, which extends the accuracy of the resummed results
  to power suppressed terms 
  (at the leading logarithmic level). As a consequence, the SCET result is
  the unimproved result, while the dQCD result is closer to the
  collinear-improved result. We will come back to an assessment of 
the impact of this
  term in the context of the SCET vs. dQCD comparison in
  Sect.~\ref{sec:pheno} below.

We have thus shown that the dQCD result coincides with the SCET result
in two steps. In the first step, we have changed the resummed scale
from $\frac{m_H^2}{\bar N^2}$ to $\mu^2_s$. If
$\ln\frac{m_H^2}{\mu_s^2\bar N^2}$ is not a large log, this leads to
the intermediate result eq.~(\ref{Cqcdmus}), which only differs from
the starting dQCD expression by a factor ${\bm C}^{(1)}_r(N,m_H^2,
\mu_s^2)$, that only contains subleading logarithmic terms, generated
by the interference of the neglected $\Ord(\as^3)$ terms with the logs
which are being resummed.  In the second step, the result
eq.~(\ref{Cqcdmus}) is found to coincide with the exact Mellin
transform of the SCET expression eq.~(\ref{eq:exactmt}), up to
power-suppressed terms: hence the SCET and QCD expressions differ by a
combined factor ${\bm C}^{(2)}_r(N,m_H^2, \mu_s^2)$ which contains
both logarithmically subleading and power-suppressed terms.

We conclude this section by briefly discussing the relation of the
result we just obtained to the ``master formula'' relating dQCD and
SCET given in ref.~\cite{Bonvini:2013td}. In that reference, we
started with the SCET expression, and we rewrote it by freely
modifying subleading terms (both log and power suppressed). We ended up
with the 
expression of 
eq.~(\ref{eq:dqcdtf}), but with only the $A_1$ and $A_2$
contributions to $A(\as)$ included in  $S^{\rm\scriptscriptstyle
  dQCD}\left(\mu_s^2,\frac{m_H^2}{\bar N^2} \right)$. Because however  
$S^{\rm\scriptscriptstyle
  dQCD}\left(m_H^2,\mu_s^2\right)$ also includes the $A_3$ term, if
one substitutes eq.~(\ref{Sqcdsplit}) with the two terms on the
right-hand side computed thus in the expression eq.~(\ref{eq:cqcdhat})
of the dQCD coefficient, one gets a result which differs from the
starting dQCD coefficient function by a multiplicative factor
\be\label{eq:cr0def}
{\bm C}^{(0)}_r(N,m_H^2, \mu_s^2)=
\exp\int_{\mu_s^2}^{\frac{m_H^2}{\bar N^2}} \frac{d \mu^2}{\mu^2}\,
\ln \frac{m_H^2}{\mu^2 \bar N^2}
\left[A(\as(\mu^2))- A_1 \as(\mu^2) - A_2 \as^2 (\mu^2) \right].
\ee
This is the result of ref.~\cite{Bonvini:2013td} (where ${\bm
  C}^{(0)}_r(N,m_H^2, \mu_s^2)$ was called $C_r$), which thus
corresponds to the very first step of the derivation presented here.


\section{Factorization and saddle-point approximation}
\label{sec:saddle}

The main result of the previous  Section is that the Landau pole can be
removed from the dQCD expression, at the cost of introducing the
dependence on an extra scale $\mu_s$. The ensuing expression
eq.~(\ref{Cqcdmus1}) only differs from the SCET result by
logarithmically subleading and power-suppressed terms.

However, one must then make a choice for the scale $\mu_s$, such that
the Landau pole is avoided. If, for instance, one chooses $\mu_s$
proportional to the partonic scale $m_H(1-z)$, with $z$ the parton
momentum fraction, (a similar choice is made in
ref.~\cite{Beneke:2011mq} for threshold resummation of the top
production cross-section) then the Landau pole reappears when
integrating over $z$. In
refs.~\cite{Becher:2006nr,Becher:2006mr,Becher:2007ty,Ahrens:2008qu}
it was suggested that one may choose $\mu_s$ to be proportional to a
hadronic scale:
\be
\mu_s =m_H(1-\tau)g(\tau),
\label{eq:BN}
\ee
where the function $g(\tau)$ is fixed in two different ways, both of
which attempt to minimize the contribution to the cross-section coming
from the one-loop term in  $\tilde
s_\textrm{Higgs}$ eq.~(\ref{softfunction}). As discussed in
Sect.~\ref{sec:analytical}, this choice is justified to the extent
that the
condition eq.~(\ref{eq:logcond}) is satisfied.
Henceforth, unless otherwise stated, SCET results will be presented by
taking for $\mu_s$ the average of these scale choices. A 
discussion of scale choices will then be provided in the end 
Sect.~\ref{sec:pheno} (in particular figure~\ref{fig:muvar1}) below.

With this choice, the problem of the
Landau pole does not arise; however, the partonic cross-section
depends on a hadronic scale, and thus, because $C$
depends on $\tau$,
the hadronic cross-section 
eq.~(\ref{sigmahad}) no longer has the form of a convolution. 
In particular, this means that the hadronic
cross-section no longer
factorizes upon taking a Mellin transform eq.~(\ref{eq:mellin}) --- we
will refer to this as breaking of Mellin factorization, or
factorization breaking, for short. 

A priori, this appears to prevent a direct comparison of the SCET and
dQCD expressions at the level of partonic cross-sections: the
comparison can only be done at the hadronic level. Indeed, the scale
whose logs are  being resummed in eq.~(\ref{Cqcdmus1}) is $\mu_s$; it
follows that, as already pointed out in ref.~\cite{Bonvini:2012az}, 
the equivalence of the dQCD and SCET resummation only holds if no
further logs are generated by the convolution integral, which in turn
depends on the form of the parton distributions. Quite apart from this
issue of principle, there remains the practical issue that,
apparently, a comparison of
resummed expressions obtained with this choice of scale to
the standard factorized expressions, is only possible once
a particular PDF has been chosen.

We will now show that this difficulty can be circumvented by using a
saddle-point method. Before doing this, we dispose of a technical
difficulty.  As mentioned in the previous section, because of the
Landau pole, $\CqcdM(N,m_H^2)$ eq.~\eqref{eq:cqcdhat} does not have a
Mellin inverse; however, any finite-order truncation of it does. There
are several ways of dealing with this problem which (unlike the method
of the previous Section) do not require introducing an extra
scale. They all amount to expanding the resummed result in powers of
$\as$, performing the Mellin inversion term by term, and viewing the
divergent series which is obtained as an asymptotic series. As shown
in ref.~\cite{Forte:2006mi}, one has to go extremely close to the 
Landau pole for this to make any
difference: it is only for $\tau\gsim 0.9$ that resummed results
obtained summing the asymptotic series in different ways  differ by
a significant amount. Hence, for the sake of the discussion in the present
paper this point is immaterial, especially since we mostly deal with
processes for which the resummed series is perturbative,
i.e.\ $\as\ln^2(1-\tau)\ll 1$.

In all the  subsequent discussion the Mellin inversion integral 
\be
\sigqcd(\tau,m_H^2)=\frac{1}{2 \pi i} \int_{c-i \infty}^{c + i \infty}
dN\, \tau^{-N} \CqcdM(N,m_H^2)\LumM(N)
\label{inversemellin}
\ee
can be thought of as being performed while taking for
$\CqcdM(N,m_H^2)$ any sufficiently high-order truncation of the
perturbative expansion. In practice, the integrals will be computed by
choosing a path which intercepts the real axis to the left of the Landau pole,
but to the right of of all other singularities of the Mellin transform
(this is the so-called minimal prescription~\cite{Catani:1996yz}).

We now explain how using the saddle-point method we can compare
results obtained using  the scale choice eq.~(\ref{eq:BN}) with standard
factorized results at the level of partonic cross-sections: the idea is
that the saddle-point approximation gives a way of directly relating
hadronic and partonic kinematics~\cite{Bonvini:2010tp,Bonvini:2012an}.
In the saddle-point approximation, the Mellin inversion integral
eq.~(\ref{inversemellin}) is given by
\be\label{saddleapp}
\sigqcd(\tau, m_H^2)\approx Z^{\rm\scriptscriptstyle dQCD}\tau^{-\nzdqcd(\tau)}
\LumM(\nzdqcd(\tau)) \CqcdM(\nzdqcd(\tau),m_H^2)
\ee
where the saddle point $\nzdqcd(\tau)$ is found by solving for the condition
\be\label{saddlecond}
\frac{d}{dN}E_{\rm\scriptscriptstyle dQCD}(\tau,N;m_H^2)\Bigg|_{N=\nzdqcd}=0
\ee
where 
\be\label{edef}
E_{\rm\scriptscriptstyle dQCD}(\tau, N; m_H^2)
=N\ln\frac{1}{\tau}+\ln\CqcdM(N,m_H^2)+\ln\LumM(N)
\ee 
and $Z^{\rm\scriptscriptstyle dQCD}$ is a fluctuation term  given by
\be
\label{zdef}
Z^{\rm\scriptscriptstyle dQCD}
=\frac{1}{\sqrt{2\pi E''_{\rm\scriptscriptstyle dQCD}(\tau,\nzdqcd(\tau);m_H^2)}}.
\ee

The main reason why the saddle-point expression eq.~(\ref{saddleapp})
is interesting is that it turns the convolution in the $\tau$-space expression
eq.~(\ref{sigmahad}) into an ordinary product, if   the
$\tau$-space expression  is viewed
as  an inverse Mellin transform, with  the Mellin inversion
integral evaluated  by saddle point. 
Indeed, each factor in eq.~(\ref{saddleapp})
is a function of $\tau$ through $\nzdqcd(\tau)$. A further simplification
comes from the fact that in practice the sub-asymptotic
fluctuation term 
$Z^{\rm\scriptscriptstyle dQCD}$
eq.~(\ref{zdef})  
is dominated by the parton luminosity, so that it is essentially
independent of the coefficient function. 

This provides us with an elegant solution to the problem of
comparing momentum-space hadron-level cross-sections in a way which separates 
the effect of the parton luminosity and of the coefficient function.
Indeed, consider the case of the SCET resummed coefficient function
eq.~(\ref{eq:resummdyscetonescale}). 
If one uses such an expression in the expression
eq.~(\ref{sigmahad}) of the hadronic cross section, it is still true
that
\be\label{eq:mellinmus}
\boldsymbol{\sigma}(N,m_H^2,\mu_s^2) = \CscetM(N,m_H^2,\mu_s^2)\LumM(N),
\ee
provided only the Mellin transform is performed at fixed $\mu_s$. Of
course, if $\mu_s$ depends on $\tau$, then upon Mellin inversion
$\boldsymbol{\sigma}(N,m_H^2,\mu_s^2)$ 
is not the full Mellin transform of
$\boldsymbol{\sigma}(\tau,m_H^2,\mu_s)$, in that the $\tau$-dependence
through $\mu_s$ is not transformed.

Be that as it may, it is still true that 
\be
\sigscet(\tau, m_H^2,\mu_s^2) \approx
 Z^{\rm\scriptscriptstyle SCET}
\tau^{-\nzscet(\tau)}
\LumM(\nzscet(\tau), m_H^2) \CscetM(\nzscet(\tau),m_H^2,\mu_s^2),
\label{sigmascetapprox}
\ee
where now
\be\label{edefscet}
E_{\rm\scriptscriptstyle SCET}(\tau,N;m_H^2)
=N\ln\frac{1}{\tau}+\ln\CscetM(N,m_H^2)+\ln\LumM(N),
\ee
and the position of the saddle point and the fluctuation term are
determined in analogy to eqs.~(\ref{saddlecond},\ref{zdef}).

Therefore, when comparing  the $\tau$-space  dQCD and SCET
expressions, we can separate the effect of the PDF from that of the
coefficient function. The former is entirely due to the fact
that, in general, the position of the saddle point is different,
so that $\nzscet\not=\nzdqcd$, and thus 
the PDF-dependent factors $\LumM$ and $Z$ do not cancel in the ratio
of $\sigscet(\tau, m_H^2,\mu_s^2)$ to $\sigqcd(\tau, m_H^2)$ 
because they are evaluated at different values of $N_0(\tau)$.
The latter, instead, is due to the fact that even at the same value
of $N$, $\CqcdM$ and $\CscetM$ do not coincide. The violation of
Mellin factorization now  manifests itself as the fact that, while
$\nzdqcd$  only depends on $\tau$ through the solution of the
saddle point condition eq.~(\ref{saddlecond}), $\nzscet$ has a
further, explicit $\tau$ dependence, due to the 
fact that  $\mu_s$ now also
  depends on $\tau$.

\begin{figure}[t]
\centering
{\includegraphics[width=.75\columnwidth]
{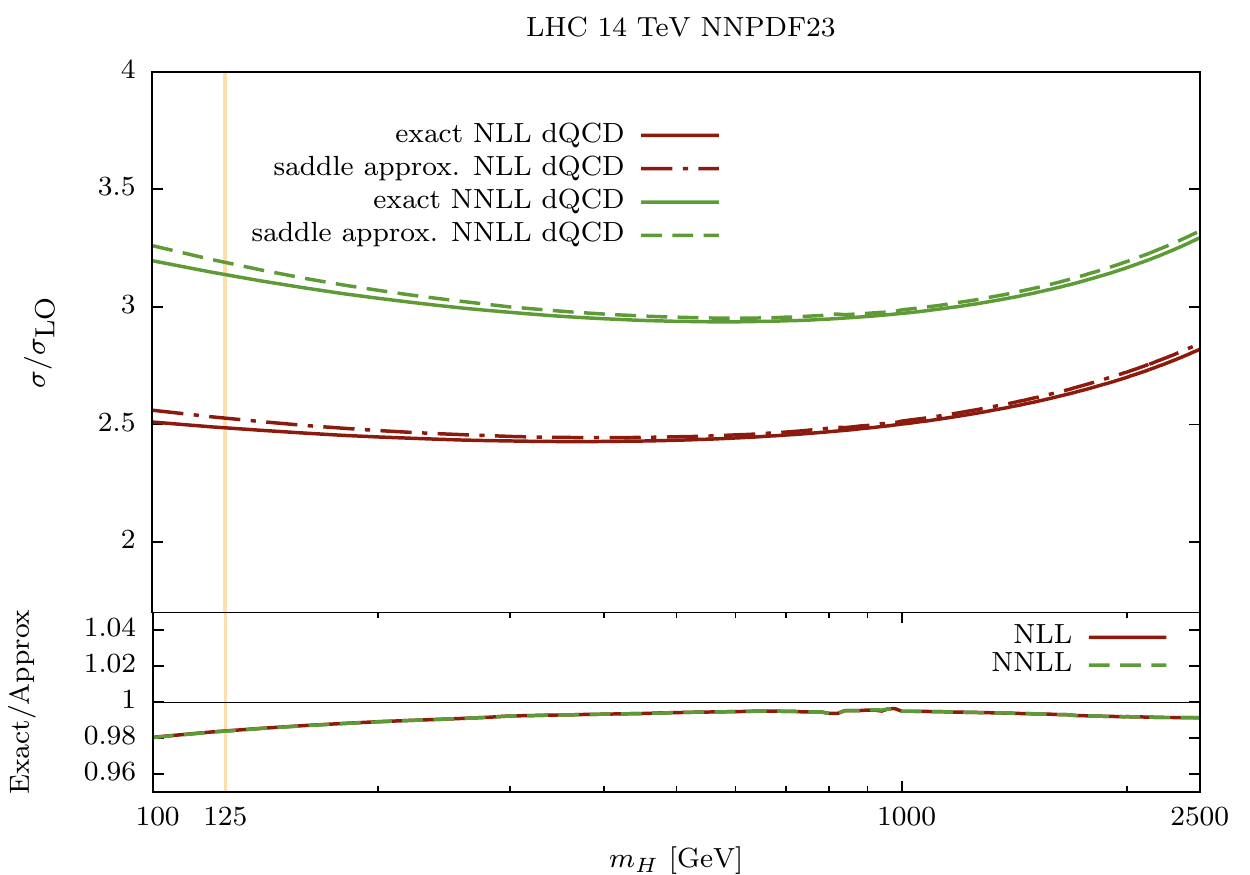}}
{\includegraphics[width=.75\columnwidth]
{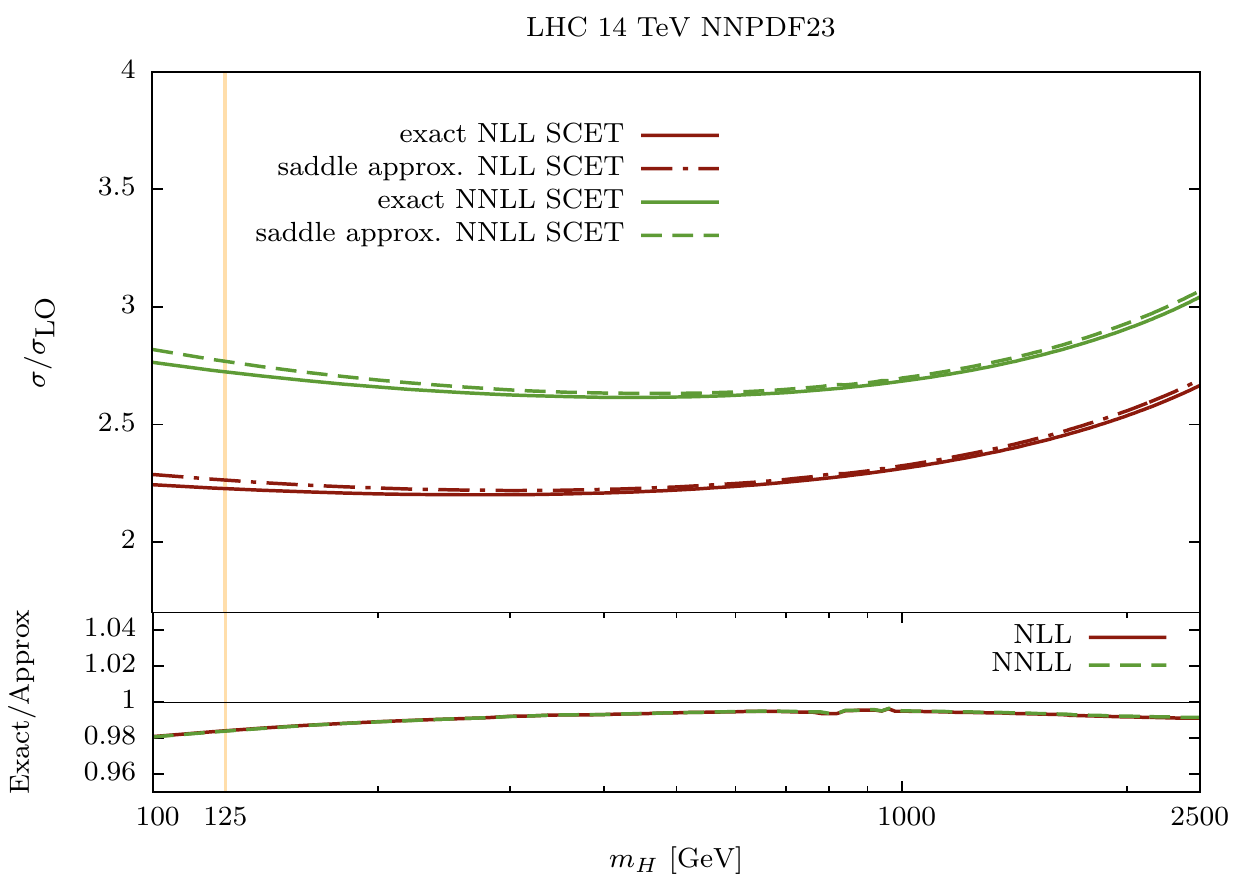}}
\caption{Comparison of the saddle-point approximation (dashed curves) 
to the exact
  result (solid curves) 
for the ratio of the resummed to leading-order cross section for
  Higgs in gluon fusion at NLL (bottom curves, green) and NNLL (top
  curves, red), computed using dQCD (top
  plot) and SCET (bottom plot), at LHC  $\sqrt{s} = 14$ TeV, for a wide
  (unphysical) range of the ``Higgs'' mass. We use NNPDF2.3 PDFs with 
$\as(M_Z^2)=0.118$. 
}\label{fig:saddlepointapprox}
\end{figure}

We can now use this method to compare the dQCD resummation to the SCET
resummation with the scale choice eq.~(\ref{eq:BN}). 
Here and henceforth
all
dQCD results
are obtained using a variant of the {\tt ResHiggs} code~\cite{gghiggs} of
ref.~\cite{Bonvini:2014joa}; SCET results are
obtained using our own computer implementation, which has been cross-checked and
benchmarked against the {\tt RGHiggs} code~\cite{rghiggs} of
refs.~\cite{Ahrens:2008qu,Ahrens:2008nc,Ahrens:2010rs}.
First, we check
the accuracy of the saddle approximation. In 
figure~\ref{fig:saddlepointapprox} we compare the resummed coefficient
function in dQCD and SCET to its saddle approximation, for a wide
range of Higgs masses. In the case of SCET, $\mu_s$ is chosen as
suggested in refs.~\cite{Becher:2006nr,Becher:2007ty,Ahrens:2008nc},
i.e.\ according to  eq.~\eqref{eq:BN} and with the function $g(\tau)$
as given there (see Section~\ref{sec:pheno} and specifically
figure~\ref{fig:crallmh} below 
for a more explicit discussion). Here and in the sequel we will always
use  NNLO NNPDF2.3
PDFs~\cite{Ball:2012cx}, 
with
$\as(M_Z^2)=0.118$ and a maximum $n_f=5$ flavor scheme.   
Note that, as mentioned in the introduction,
here and henceforth we will always
consider a range of value of $m_H$ that goes well beyond the
acceptable physical Higgs mass region, in order to see the qualitative
behaviour also in high-mass regions closer to threshold, which may be
of relevance for other processes, such as Drell-Yan production.
It is clear that the relative deviation between the exact result and
the saddle approximation is always below 2\%. In fact, note that the
deviation goes in the same direction for the SCET and dQCD results, so
using the saddle approximation to evaluate their ratio is yet more
accurate.

Having established the accuracy of the saddle-point approximation, we
proceed to estimating the violation of Mellin factorization, which, as
mentioned, can only manifest itself in the position of the saddle
being different for dQCD and SCET, $\nzscet\not=\nzdqcd$. 
The position of the saddle is determined through the condition
eq.~(\ref{saddlecond}) by the explicit form of the coefficient
function: hence, even a pair of coefficient functions which do respect
factorization, but differ in any aspect (such as the way the
resummation is implemented) will generally lead to different 
saddle-points values. Therefore the deviation from unity
of the ratio $\frac{\nzdqcd}{\nzscet}$ provides an upper bound to the
violation of factorization: the shift in the position of the saddle
measures the maximal effect of factorization violation.

\begin{figure}[htbp]
\centering
{\includegraphics[width=.75\columnwidth]
{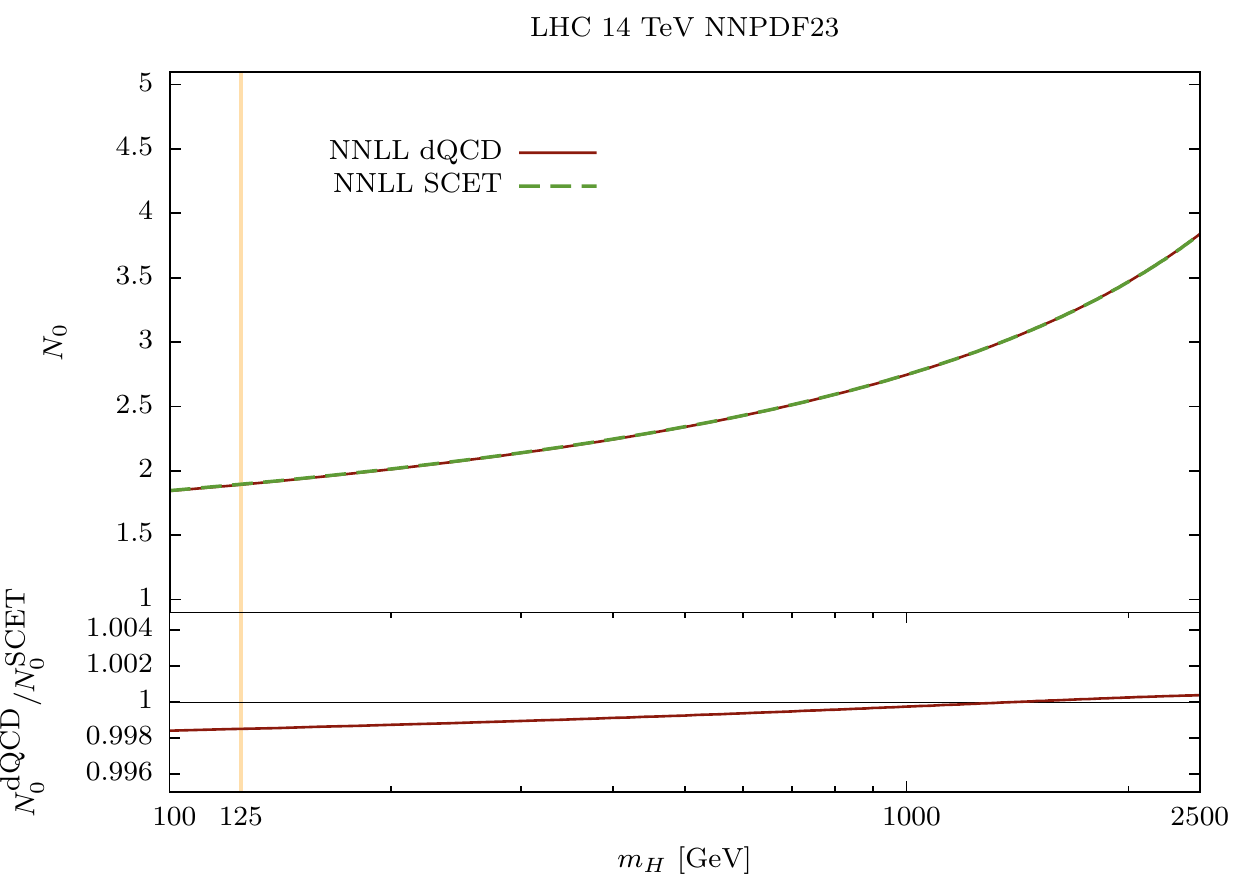}}
\caption{The saddle points  $\nzdqcd$ (solid curve) and $\nzscet$ (dashed)
  computed using respectively  eq.~(\ref{edef}) and
  eq.~(\ref{edefscet}) in  eq.~(\ref{saddlecond}) (top); their
  ratio is also shown (bottom).  All settings are the
  same as in figure~\ref{fig:saddlepointapprox}.
}\label{fig:saddlepointfact1}
\end{figure}

In figure~\ref{fig:saddlepointfact1} we display $\nzdqcd$, $\nzscet$ and
their ratio, for the same kinematics and with the same settings as for
figure~\ref{fig:saddlepointapprox}. It is clear that the difference is
at the permille level in the whole range.
We conclude that, in practice, the effect of the violation of
Mellin factorization is negligible. We understand this as a consequence of
the fact that the position of the saddle point is dominated by the
parton luminosity, i.e.\ the contribution of $\CqcdM(N,m_H^2)$ and
$\CscetM(N,m_H^2)$ to eqs.~\eqref{edef} and \eqref{edefscet}
respectively is completely negligible, as already observed in
ref.~\cite{Bonvini:2010tp} (see in particular figure~3). We conclude
that in practice $\nzdqcd\approx\nzscet$ to very good approximation.

It then immediately follows from eq.~\eqref{saddleapp} that
\be\label{eq:sigrateff}
\frac{\sigqcd(\tau,m_H^2)}
{\sigscet(\tau,m_H^2,\mu_s^2(\tau))} \approx 
\frac{\CqcdM(\nzdqcd(\tau),m_H^2)}
{\CscetM(\nzscet(\tau),m_H^2,\mu_s^2(\tau))}
={\bm C}^{(2)}_r(N_0(\tau),m_H^2,\mu_s^2(\tau)),
\ee
where the first approximate equality follows from
the fact that $\nzdqcd\approx\nzscet=N_0$
and $Z^{\rm\scriptscriptstyle dQCD}\approx Z^{\rm\scriptscriptstyle SCET}$,
in the last step we have used the definition
eq.~\eqref{eq:cr2def}, and the only remaining trace of violation of
factorization is in the the dependence of $\mu_s$ on $\tau$, which we
have explicitly indicated.
We have  checked that ${\bm C}^{(2)}_r(N_0(\tau),m_H^2,\mu_s^2)$
coincides with the ratio of physical cross-sections according to
eq.~(\ref{eq:sigrateff}), up to corrections which never exceed 0.4\%
in the region shown, in agreement with the conclusion reached from
figure~\ref{fig:saddlepointfact1}.

We conclude that with this scale choice the violation of Mellin factorization
in the SCET expression has a negligible impact, and that further
differences between the SCET and dQCD results in $\tau$ space and at
the hadronic level can be reduced to parton-level results in
$N$-space, and in particular   can be assessed by studying the
three ratio functions  ${\bm C}^{(i)}_r(N,m_H^2,\mu_s^2)$
eqs.~(\ref{eq:cr1def},\ref{eq:cr2def},\ref{eq:cr0def}), all evaluated
at $N=N_0(\tau)$.

Before concluding this section, we note that the saddle-point approach
allows us to better understand and generalize a result of
ref.~\cite{Sterman:2013nya}. There, it was argued that, if the
luminosity behaves as
\be
\Lum(z)\sim z^{-s_1},\qquad s_1>0
\label{lum}
\ee
in the relevant range of integration, as assumed for example in
ref.~\cite{Becher:2007ty}, then
\be
\sigma(\tau,m_H^2)=\int_\tau^1\frac{dz}{z}\,\Lum\left(\frac{\tau}{z}\right)
C(z,m_H^2)=\Lum(\tau)\int_\tau^1\frac{dz}{z}\,z^{s_1}C(z,m_H^2),
\ee
which is just the $s_1^{\rm th}$ Mellin moment of
$C(z,m_H^2)$ times the luminosity, up to corrections of order
$\Ord(\tau^{s_1})$, which is negligible for $\tau\ll 1$. 
The same result can be obtained through the 
saddle-point approximation. Indeed, assuming eq.~\eqref{lum},
we have
\be
\sigma(\tau,m_H^2)=\frac{1}{2\pi i}\int_{c-i\infty}^{c+i\infty} dN\,
\tau^{-N}\frac{1}{N-s_1}{\bm C}(N,m_H^2).
\ee
It is easy to check that, for $\tau\ll 1$, the coefficient function does not
affect the position of the saddle point, which is given by
\be
N_0\simeq s_1+\frac{1}{\ln\frac{1}{\tau}},
\ee
and the saddle-point approximation of the inversion integral gives
\be
\sigma(\tau,m_H^2)\approx
\frac{e}{\sqrt{2\pi}}\tau^{-s_1}{\bm C}(N_0,m_H^2)
\simeq\frac{e}{\sqrt{2\pi}}\Lum(\tau){\bm C}(s_1,m_H^2)
\ee
where in the last step we have neglected corrections of order
$1/|\ln\tau|$ to ${\bm C}(s_1,m_H^2)$, as appropriate for $\tau\ll 1$.
This is the same result obtained in ref.~\cite{Sterman:2013nya},
since $e/\sqrt{2\pi}\simeq 1.084$.

\section{Logarithmically subleading vs.\ power-suppressed differences:
  an assessment}
\label{sec:pheno}

Having established that, through the saddle-point method, SCET and dQCD
cross sections can be compared at the partonic level, we now assess
the quantitative impact of the differences that were presented
analytically in Sect.~\ref{sec:analytical}.

In that section, we saw that we can obtain the SCET expression from
the dQCD one in two steps: first, by introducing the dependence on the
scale $\mu_s$ and expanding in order to get rid of the Landau pole,
which leads to an expression which differs from the starting dQCD one
by logarithmically subleading terms. In  $N$ space,
the ratio of these two expression
is given  by ${\bm C}^{(1)}_r(N,m_H^2, \mu_s^2)$
eq.~\eqref{eq:cr1def}. In the second step, power-suppressed terms are
introduced, in order to get the standard SCET expression, whose
ratio to the starting dQCD one is ${\bm C}^{(2)}_r(N,m_H^2, \mu_s^2)$
eq.~(\ref{eq:cr2def}). A preliminary step, which corresponds to the
``master formula'' of ref.~\cite{Bonvini:2013td}, corresponds to
comparing results after introducing the dependence on $\mu_s$ in the
dQCD result, but before expanding it out, which may already
introduce some subleading $\mu_s$-dependent terms, whose ratio to the
starting QCD expression is given by ${\bm C}^{(0)}_r(N,m_H^2, \mu_s^2)$
eq.~(\ref{eq:cr0def}).
In $z$ space, in the first step the starting dQCD expression is
turned into an expression which has the same form as the SCET
result, but with the soft function eq.~\eqref{softfunction} replaced
by its large-$z$ form eq.~\eqref{eq:scetlarget}. In the second
step, one ends up with the standard resummed SCET result.

\begin{figure}[t]
\centering
{\includegraphics[width=.495\columnwidth]
{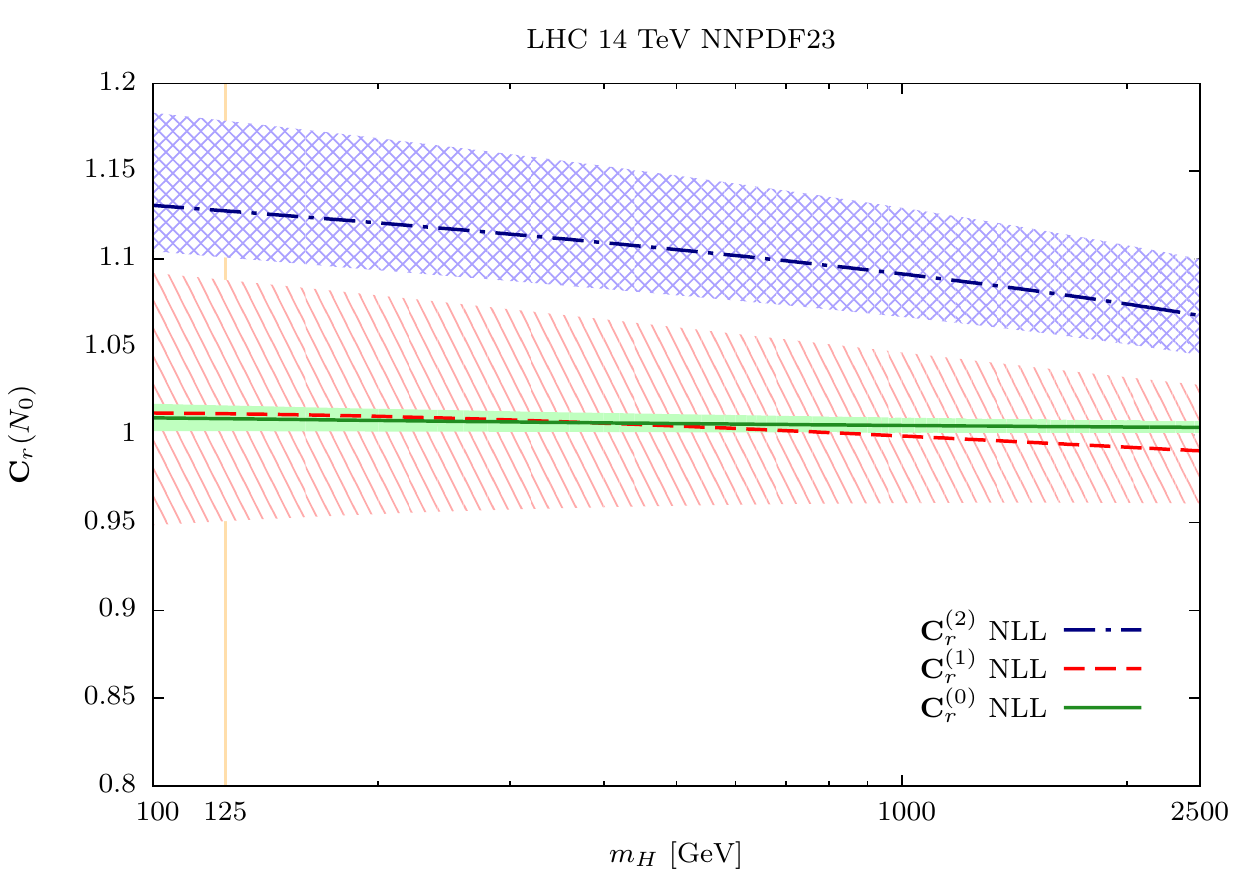}}
{\includegraphics[width=.495\columnwidth]
{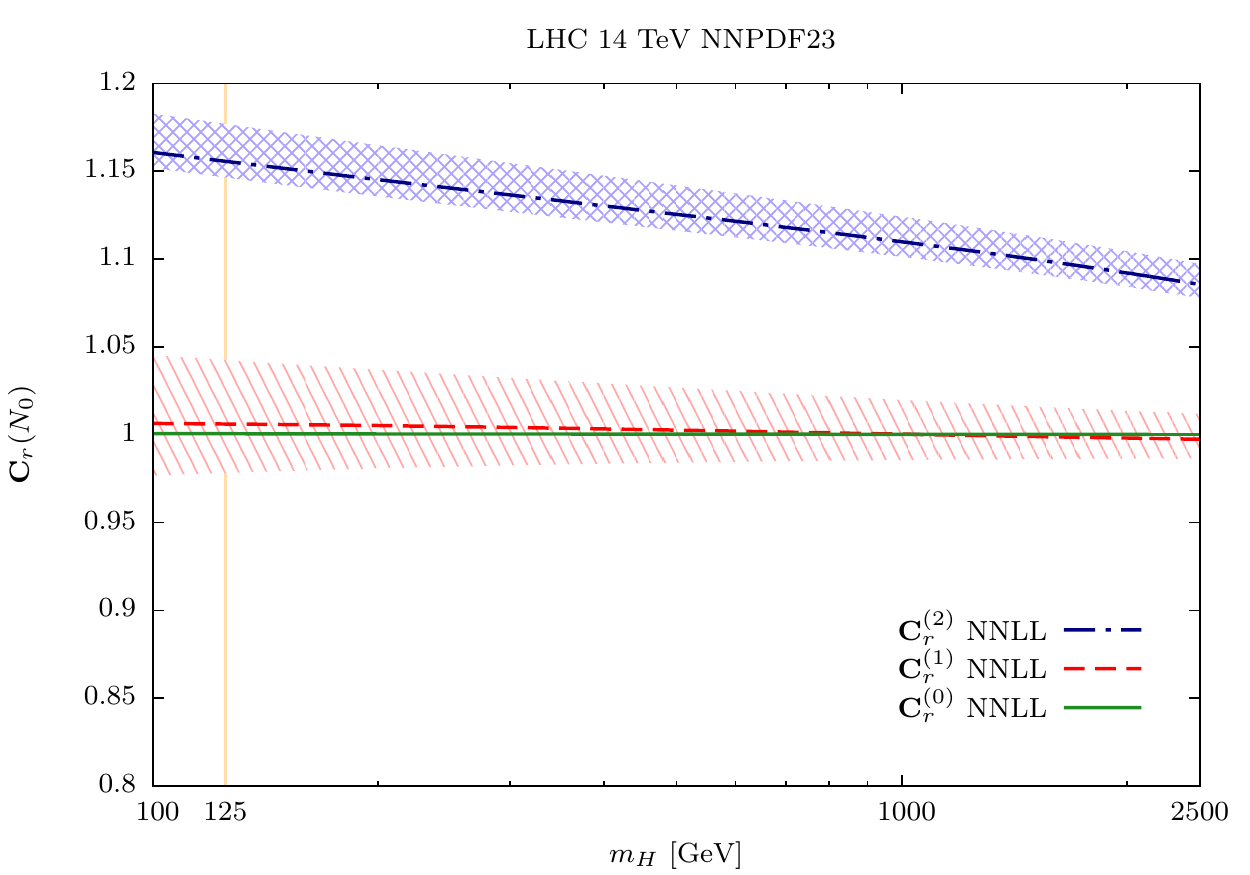}}
\caption{The ratio functions ${\bm C}^{(i)}_r(N_0(\tau),M^2,\mu_s^2)$
  eqs.~(\ref{eq:cr1def},\ref{eq:cr2def},\ref{eq:cr0def}), computed at
  NLL (left) and NNLL (right) as a function of the saddle point
  $N_0(\tau)$, and plotted versus the Higgs mass $m_H$. All settings
  are the same as in figure~\ref{fig:saddlepointapprox}. The curves
  shown are, from top to bottom, ${\bm C}^{(2)}_r$, ${\bm C}^{(0)}_r$,
  and ${\bm C}^{(1)}_r$. In each case, 
 the edges of 
  the band correspond to the two different choices for $\mu_s$ from
  refs.~\cite{Becher:2006nr,Becher:2007ty,Ahrens:2008nc},  and
  the central curve is their average.}
\label{fig:crallmh}
\end{figure}

In figure~\ref{fig:crallmh}, we plot the three ratio functions
${\bm C}^{(i)}_r(N,m_H^2,\mu_s^2)$, all evaluated (at the NLL and NNLL
level) at $N=N_0(\tau)$, as functions of $m_H$.  In each case,
we show  results as a band whose edges are obtained using $\mu_s$
eq.~(\ref{eq:BN}) 
computed using either of
the two choices for $g(\tau)$ from
  refs.~\cite{Becher:2006nr,Becher:2007ty,Ahrens:2008nc}, with a
  central curve obtained using the average of these two scale choices;
  of course, the dependence on the choice of $\mu_s$ is always rather
  less at NNLL than at NLL.

It is apparent from figure~\ref{fig:crallmh} that the difference
between SCET and dQCD resummation, measured by   ${\bm
  C}^{(2)}_r$, is sizable: at NNLO, it is of order 15\% for the
physical value of the Higgs mass, slowly decreasing as the Higgs mass
increases towards the threshold --- as one would expect, as in the
threshold limit all different forms of resummation  coincide by
construction. This difference is as large as the effect of the
resummation itself on the matched NNLO+NNLL result.
Indeed, at the matched level,
in dQCD resummation leads to an increase by about
10\% of the unresummed NNLO
result~\cite{Dittmaier:2011ti,deFlorian:2012yg}. 
Using instead the SCET prediction with the scale choice
eq.~\eqref{eq:BN}, and 
everything else being equal, we find that at the matched level
the effect of resummation on the fixed-order result  is negligible, 
at the level of
a few percent. This fact is sometimes
obscured, because in 
comparisons of SCET and dQCD different
ingredients are often included in either calculation, instead of comparing
like with like.

However, if we look at  ${\bm C}^{(1)}_r$, i.e., we tune all power
suppressed terms in the SCET
result so that it only differs by logarithmically subleading terms
from the dQCD result, the difference is down to a few percent at NNLL, with
SCET just above or just below the dQCD result, according to which
choice is made for $\mu_s$. If
finally we look at ${\bm C}^{(0)}_r$, i.e.\ the difference between dQCD and its
$\mu_s$-dependent form, the difference is at the sub-percent level
with any choice of $\mu_s$.

We conclude that the difference between what we have been calling
respectively 
SCET and dQCD is large; however, it is almost entirely due to
power-suppressed terms, which are really not a feature of either dQCD
or SCET, but rather, an instance of the resummation ambiguities which
are present in any approach and which have been previously discussed
in refs.~\cite{Bonvini:2010tp,Kramer:1996iq,Ball:2013bra}. Indeed, as
we have seen in Sect.~\ref{sec:analytical}, the leading $\Ord\left((
1-z)\right)$  power difference  
between the SCET and dQCD result can
be expressed as a $\sqrt{z}$ prefactor, according to
eq.~(\ref{eq:leadpowcorr}). This prefactor (or rather, its $N$-space
transformed version) is thus mostly responsible for the deviation of ${\bm
  C}^{(2)}_r$ from one, i.e.\ for the difference between the SCET and
dQCD results. As already mentioned in Sect.~\ref{sec:analytical}, this
factor can be related to the collinear improvement discussed in
Refs.~\cite{Kramer:1996iq,Ball:2013bra}, and thus the difference is akin to that
between versions of resummation which do  or do not include this
collinear improvement.

Clearly, the size of the deviation measured by ${\bm
  C}^{(1)}_r$ will crucially depend on the choice of scale $\mu_s$.
 In
particular, if $\mu_s=m_H/\bar N$, then $F(L,\mu_s)=1$ in
eq.~(\ref{eq:fexp}).
 Now, the high
accuracy of the saddle-point approximation means that the Mellin
inversion integral eq.~(\ref{inversemellin}) is completely dominated
by $N=N_0(\tau)$. It follows that 
\be
\label{optscale}
\mu_s=\frac{m_H}{\bar N_0(\tau)}
\ee
is the optimal choice of scale if one wishes to remove the Landau pole
from the dQCD resummed expression by replacing it with
eq.~(\ref{Cqcdmus1}), while leaving the result unchanged  as much as
possible. 
Equivalently, this is the choice that minimizes
$\abs{\ln\frac{m_H^2}{\mu_s^2\bar N^2}}$, and thus maximizes the
inequality eq.~(\ref{eq:logcond}) which justifies this treatment of
the Landau pole. 

Also, this provides us with a criterion to judge whether any other
(hadronic) choice of $\mu_s$ will lead to large or small deviation from
the starting expressions. In
refs.~\cite{Becher:2006nr,Becher:2006mr,Becher:2007ty,Ahrens:2008qu}
the function $g(\tau)$ which enters the scale choice eq.~(\ref{eq:BN})
is chosen as an ad-hoc functional form, whose parameters are
determined, as already mentioned,
by demanding that the
one-loop contribution of $\tilde s_\textrm{Higgs}$ 
to the cross-section be minimized.

\begin{figure}[t]
\centering
{\includegraphics[width=.75\columnwidth]{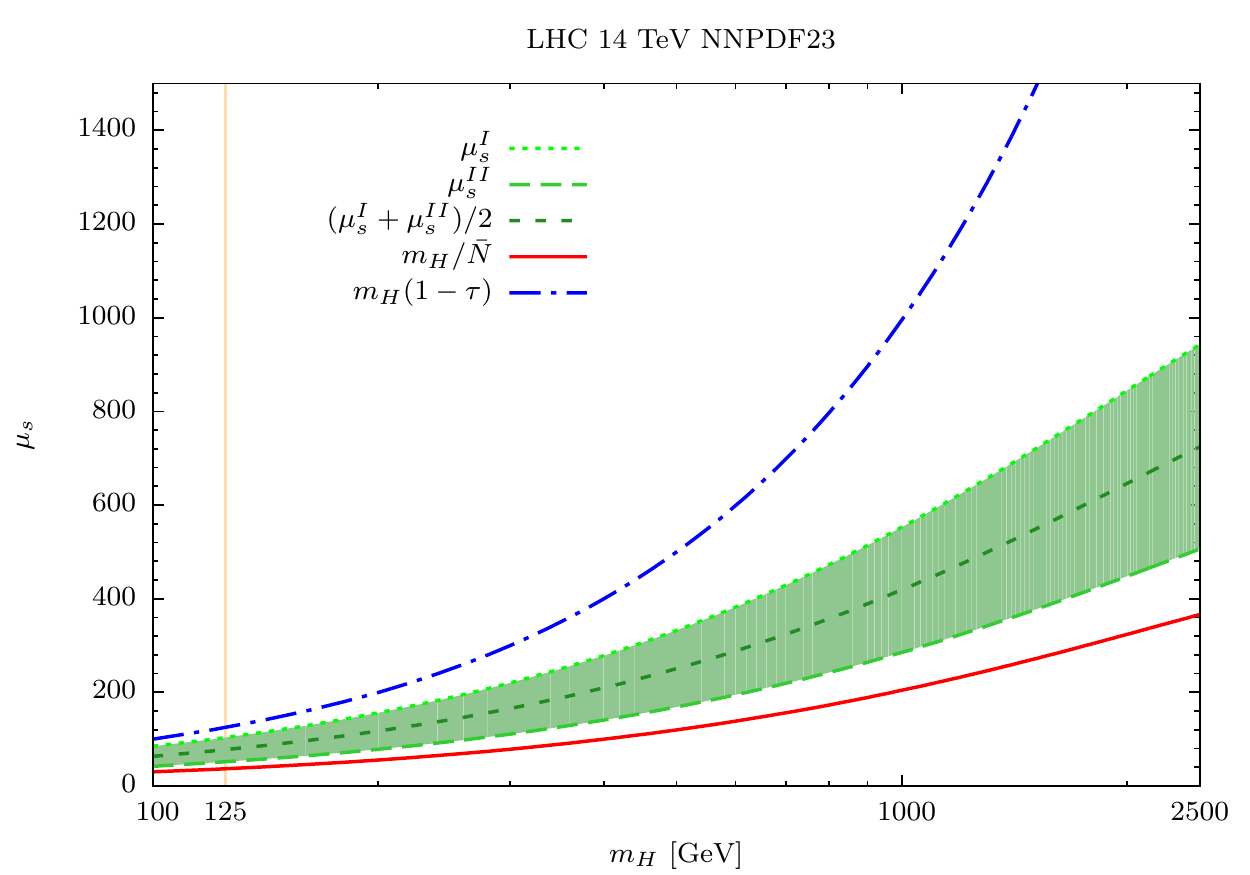}}
\caption{Comparison of choices of of soft scale $\mu_s$: from bottom
  to top, the ``optimal'' scale choice eq.~\eqref{optscale};
  eq.~(\ref{eq:BN}) with $g(\tau)$ determined as in
  refs.~\cite{Becher:2006nr,Becher:2007ty,Ahrens:2008nc}; and the
  naive hadronic scale choice $\mu_s=m_H(1-\tau)$.}
\label{fig:muvar1}
\end{figure}

It is thus interesting to compare the choice of scale
eq.~\eqref{eq:BN}, with  $g(\tau)$ as in
refs.~\cite{Becher:2006nr,Becher:2006mr,Becher:2007ty,Ahrens:2008qu},
with the ``optimal'' choice eq.~(\ref{optscale}). We do this in
figure~\ref{fig:muvar1}, where, for reference, the naive hadronic scale
choice which corresponds of taking $g(\tau)=1$ in eq.~(\ref{eq:BN}) is
also shown. As already mentioned, in
refs.~\cite{Becher:2006nr,Becher:2006mr,Becher:2007ty,Ahrens:2008qu}
two different ways to actually determine  $g(\tau)$ are suggested:
either 
by choosing the absolute minimum of the one-loop correction
(referred to as $\mu_s^{II}$ in figure~\ref{fig:muvar1}), or
starting from a high scale, and choosing $\mu_s$ as the value ($\mu_s^I$)
at which
the one-loop term drops below 15\%. In either case, an empirical
parametrization which approximately implements the condition is
chosen. In figure~\ref{fig:muvar1} we show both  $\mu_s^{I}$,
$\mu_s^{II}$, and also their average.

It is clear that the scale choice eq.~(\ref{eq:BN}) and the
``optimal'' choice eq.~(\ref{optscale}) are quite close, and rather
different from the naive choice $\mu_s=m_H(1-\tau)$, which, of course,
almost coincides with $m_H$ even for $m_H$ as large as 1~TeV. This
explains the smallness of ${\bm C}^{(1)}_r$ seen in
figure~\ref{fig:crallmh}, and also (since $m_H/\mu_s\sim 2$) why the
resummation does have a visible effect, albeit not a huge one.  The
reason why these two scale choices are close is obvious: as we have
seen, the optimal scale choice leads to the vanishing of all higher
order contributions to $F(L,\mu_s)$. But $F(L,\mu_s)$ and $\tilde
s_\textrm{Higgs}$ are closely related, see eq.~\eqref{stildef}: hence
the condition of
refs.~\cite{Becher:2006nr,Becher:2006mr,Becher:2007ty,Ahrens:2008qu}
is essentially an empirical version of the more quantitative condition
eq.~(\ref{optscale}).

Before concluding, we would like to note that the actual
implementation of resummation in dQCD ref.~\cite{deFlorian:2012yg} and
SCET refs.~\cite{Becher:2006nr,Becher:2007ty,Ahrens:2008nc} differ in
many further aspects.  First of all, somewhat confusingly, a different
nomenclature is used in the literature for the logarithmic accuracies
of table~\ref{tab:count}: in particular, in
refs.~\cite{Becher:2006nr,Becher:2007ty,Ahrens:2008nc} only the
starred NLL*, NNLL*, etc., orders of resummation are considered, but
they are referred to as NLL, NNLL etc.  Specifically, the
highest-order resummation considered in that reference is NNNLL*, which
requires guessing the currently unknown $A_4$ coefficient (in
refs.~\cite{Becher:2006nr,Becher:2007ty,Ahrens:2008nc} this is done by
means of a Pad\'e approximation). However, as shown in
figure~\ref{fig:starvsnonstar}, 
NNNLL* resummation is almost indistinguishable from
NNLL, so the extra effort of upgrading from NNLL to NNNLL* seems
unwarranted, as  results actually obtained using the {\tt
  RGHiggs} code~\cite{rghiggs} at NNNLL (which is really NNNLL*) are
very close to the NNLL, discussed here and shown in all plots of this
paper.

Furthermore, the SCET resummation in
refs.~\cite{Becher:2006nr,Becher:2007ty,Ahrens:2008nc} also includes
the exponentiation of a class of constant contributions proportional
to $\pi^2$; the dQCD resummation also includes finite top and bottom
mass effects~\cite{deFlorian:2012yg}, and so on. Some of these
contributions (for instance, the $\pi^2$ exponentiation) have an
effect which is quantitatively comparable to the overall effect of the
resummation, and thus their inclusion can significantly affect
results. They were all quantitatively assessed in the respective
references, so there is no reason for us to discuss them here. Rather,
by making sure that everything else is treated in the same way, we
have focused on the effect due to the different treatment of
logarithmically subleading and power-suppressed contributions.

\section{Summary and outlook}
\label{sec:conclusions}

We have provided a careful, step-by-step comparison of SCET and dQCD
resummation, concentrating on differences in treatment of
logarithmically subleading and power suppressed contributions.
We have chosen Higgs production as a case study, but our results also
hold for related processes such as Drell-Yan production. Also, we have
considered a wide (unphysical) range of the Higgs mass, which extends
to final states with a mass of several TeV.

Our main results can be summarized as follows (not necessarily in the
order in which we have obtained them):

\begin{itemize}
\item The Landau pole can be removed from the dQCD
  resummed expression by introducing a dependence on an extra scale
  $\mu_s$. 
  This scale may be chosen in an optimal way, such that the dependence
  on it is minimized, by using a saddle-point argument.
\item The form of dQCD resummation thus obtained  only differs from
  the SCET resummed result by logarithmically subleading and by power
  suppressed terms.
\item A saddle-point argument allows for a comparison at the level of
  partonic cross-sections of different resummation schemes, even if
  the soft scale which is being resummed is chosen as a hadronic
  scale. 
\item Using this saddle-point argument, it can be shown that 
   the fact that 
  in the SCET result of
refs.~\cite{Becher:2006nr,Becher:2006mr,Becher:2007ty,Ahrens:2008nc}
the soft scale $\mu_s$ depends on a hadronic scale (and thus the
hadronic cross-section does not factorize upon Mellin transformation)
 has a negligibly small impact.
\item The logarithmically subleading differences between the SCET and
  dQCD result are small if the scale is chosen according to the
  procedure suggested in
  refs.~\cite{Becher:2006nr,Becher:2007ty,Ahrens:2008nc}, which
  appears to be an empirically motivated version of the optimal dQCD
  scale choice referred to above.
\item The power-suppressed differences between the SCET and dQCD
  result are large: in fact large enough that while NNLL resummation
  in dQCD leads to an increase of order of $\sim 10\%$ of the NNLO
  cross-section when matched to it, 
in SCET it leads to an almost negligible increase, at
  the  percent level. 
\end{itemize}

The fact that the large difference between SCET
and dQCD resummation is  dominated by power-suppressed terms means
that, as mentioned in the introduction, the difference is really
unrelated to the use of dQCD vs. SCET,  and rather to
known~\cite{Bonvini:2010tp,Kramer:1996iq,Ball:2013bra} 
ambiguities when resummation is used to improve fixed-order
computations away from threshold.

This might suggest the pessimistic conclusion that resummation, and more
generally the soft approximation, do not provide reliable results 
in this kinematic region. This is of course a possible 
point of view, which, in particular, was recently advocated in
ref.~\cite{Anastasiou:2014vaa}. However we do not think that this is
necessarily the case: rather, reliable results may
be obtained, provided 
the choice of power-suppressed contribution is optimized by
imposing requirements such as Mellin-space
analiticity~\cite{Ball:2013bra}, which can be validated by comparison
to the orders (up to NNLO) for which the full result is known.
It is interesting to observe that if this is done, it turns
out~\cite{Bonvini:2014joa} 
that the effect of resummation on Higgs
production in gluon fusion is in fact a further enhancement, which
roughly doubles the enhancement found in the ``standard'' dQCD
approach of ref.~\cite{deFlorian:2012yg}.
However, if this is  not done the conclusion that the spread of resummed
results is bigger than the impact of the resummation seems inevitable
not only for Higgs production but also for processes which are much
closer to threshold, such as, say production of a $Z^\prime$ with the
mass of a few TeV at the LHC.
\medskip

\acknowledgments

We thank G.~Ferrera for asking a question which is answered by the
argument presented in Sect.~\ref{sec:analytical}, and S.~Carrazza for
help with NNPDF parton distributions. We thank the anonymous referee
for very useful suggestions for the improvement of the paper. 
SF and GR are supported in part
by an Italian PRIN2010 grant, and SF also by a European Investment
Bank EIBURS grant, and by the European Commission through the
HiggsTools Initial Training Network PITN-GA-2012-316704.

\end{document}